\documentclass[aps,prl,showpacs,reprint,groupedaddress]{revtex4-1}

\usepackage{graphicx}
\usepackage{color} 
\usepackage{ulem}

\begin{document}


\title{Observation of Self-binding in Monolayer $^3$He}


\author{D. Sato, K. Naruse, T. Matsui and Hiroshi Fukuyama}
\email[]{hiroshi@phys.s.u-tokyo.ac.jp}
\affiliation{Department of Physics, Graduate School of Science, The University of Tokyo, 7-3-1 Hongo, Bunkyo-ku, Tokyo 113-0033, Japan}


\date{\today}

\begin{abstract}

We report clear experimental signatures of the theoretically unexpected gas-liquid transition 
in the first three monolayers of $^3$He adsorbed on graphite. 
The transition is inferred from the linear density dependence of the $\gamma$-coefficient of 
the heat capacity measured in the degenerate region (2 $\le T \le$80~mK) below a critical liquid 
density ($\rho_{c0}$). 
Surprisingly, the measured $\rho_{c0}$ values (0.6$\sim$0.9~nm$^{-2}$) are nearly the same for 
all these monolayers in spite of their quite different environments. We conclude that the 
ground-state of $^3$He in strict two dimensions is not a dilute quantum gas but a self-bound 
quantum liquid with the lowest density ever found.   

\end{abstract}


\maketitle{}



Matter can in principle be in either gas or a liquid phase at absolute zero if the quantum parameter, the zero-point kinetic energy divided by the potential energy, is large enough. The two-dimensional (2D) helium-3 ($^3$He) system has long been thought 
as the only material which stays gaseous at the ground state \cite{Note}. 
This system is experimentally realized in $^3$He monolayers adsorbed on an atomically flat and strongly attractive graphite surface. 
Most previous theories based on the variational calculations \cite{Novaco(1975),Miller(1978),Krishnamachari(1999)}, the diffusion Monte Carlo calculation \cite{Grau(2002)} and the Fermi hypernetted chain method \cite{Um(1997)} support the 
absence of self-binding of $^3$He in 2D. 
Indeed, no signature of the gas-liquid (G-L) transition was experimentally observed in the first and second layer $^3$He on graphite down to $T \approx$ 3~mK and to areal density $\rho$ = 1~nm$^{-2}$ \cite{Greywal(1990)}. 
This is in sharp contrast to monolayer $^4$He with smaller quantum parameter on graphite. 
It is well established experimentally \cite{Greywall(1993)} and theoretically \cite{Pierce(2000)} that in this system the G-L transition takes place at temperatures below 1 K and the self-bound liquid density at $T = 0$ ($\rho_{c0}$) is 4~nm$^{-2}$.

The first experimental address to this problem was made by Bhattacharyya and Gasparini \cite{Bhattacharyya(1985)}, who found a kink or small discontinuity near 100~mK in the heat capacity ($C$) of submonolayer $^3$He floated on a thin superfluid $^4$He film adsorbed on a Nuclepore substrate.
They attributed this to a puddle formation of $^3$He in 2D.
It is to be noted, however, that in this system the indirect $^3$He-$^3$He interaction mediated by ripplons in the underlying $^4$He film, which is not considered in most theoretical works, might be important.
In addition, Nuclepore is believed to be a much less uniform substrate than graphite.

Recently, Sato et al.~\cite{Sato(2010)} found the G-L transition with $\rho_{c0} \approx$ 1~nm$^{-2}$ in the heat-capacity measurements on the third layer of $^3$He on graphite down to $T=$ 1~mK.
This was inferred from a linear $\rho-$dependence of $\gamma$, the coefficient of the leading $T$-linear term of $C$ in the degenerate region, as well as a kink at $\gamma \approx \gamma_{\text{ideal}}$.
Here $\gamma _{\text{ideal}} (= \pi k_{B}^2 Am/3\hbar^2)$ is the $\gamma$ value 
of an ideal Fermi gas spreading over the whole surface area ($A$) of the substrate, and $m$ is the bare mass of $^3$He. 
Note that $\gamma$ depends only on $A$ and $m$ not on the number of particles in the 2D case.
One possible explanation for their result, which contradicts existing theory, is that, in the third layer, the relatively large plane-normal motion may stabilize the liquid phase (the quasi two-dimensionality). 
A variational calculation \cite{Brami(1994)} supports this scenario but the subsequent ones \cite{Krishnamachari(1999), Grau(2002)} do not. 
This hypothesis can be tested by extending their $C$ measurement to the first or second layer in which the substrate confinement potential is much deeper. 
The other issue is the role of surface heterogeneities in Grafoil \cite{GTIHI}, an exfoliated graphite substrate, used in most of the previous experiments including Ref.~\cite{Sato(2010)}. 
This substrate is known to have a platelet (micro-crystallite) structure with a mosaic angle spread of about 30 degrees  \cite{Takayoshi(2010)} and a platelet size of 10$\sim$100~nm \cite{Niimi(2006)}.
The role can be checked, for instance, by comparing results on the first layer of $^3$He, which is directly on the Grafoil, and those in the upper layers. 

In this Letter, we report a result of new heat-capacity measurements of three different $^3$He 
monolayers, i.e., the first, second, and third layers of $^3$He on graphite, at very low densities never explored before, using the same experimental setup as in Ref.~\cite{Sato(2010)}. 
We could determine the substrate heterogeneity effect explicitly in the first-layer measurement. 
By preplating the first layer with nonmagnetic $^4$He, the effect can thoroughly be removed in the second-layer measurement. Surprisingly, all the three monolayers show the G-L transitions with approximately the same $\rho_{c0}$ values ($0.6\sim0.9$~nm$^{-2}$). 
This indicates that the quantum gas phase is not the ground sate of $^3$He in strictly 2D and gives rise to a challenge for current many-body theories. 

\begin{figure}
\includegraphics[width=0.46\textwidth]{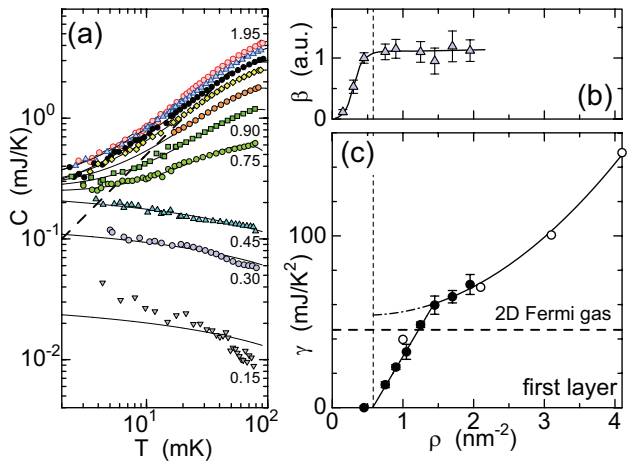}
\caption{
(a) Heat capacities ($C$) of the first layer $^3$He on Grafoil. 
The numbers are densities in nm$^{-2}$, and those not denoted are 1.05, 1.25, 1.45 and 1.70 nm$^{-2}$, respectively, from bottom to top.
The solid lines are fittings to Eq. (\ref{eq1}). 
The dashed line is $C$ = $\gamma_{\text{ideal}}T$ for a degenerate Fermi gas spreading over the whole surface. 
(b) Density dependence of the fitting parameter $\beta$ in Eq. (\ref{eq1}). 
(c) Density dependence of the fitting parameter $\gamma$ in Eq. (\ref{eq1}).  
The open circles are from Ref.~\cite{GreywallBusch(1990)}.
Only after the growth of $\beta$ is saturated, $\gamma$ starts to increase linearly with $\rho$ above 0.6~nm$^{-2}$. 
The horizontal dashed line represents $\gamma$ = $\gamma_{\text{ideal}}$.
}
\label{lfig_1}
\end{figure}

In Fig.~\ref{lfig_1}(a), we show measured $C$ data for the first layer of $^3$He adsorbed directly on the Grafoil substrate with $A$ = 556~m$^2$. 
This surface area is determined from the sub-step structure in N$_2$ adsorption isotherm measurement corresponding to the $\sqrt{3} \times \sqrt{3}$ commensurate phase formation in the first layer.
Two heat-capacity contributions with distinct $T$-dependences develop successively as a 
function of $\rho$. The data at any densities above 0.3~nm$^{-2}$ can be well fitted to 
\begin{equation}\label{eq1}
   C(T, \rho) = \gamma T - \alpha T^2 + \beta C_{\text{amor}}(T)
\end{equation}                                                                     
in the $T$-range from 4 to 80~mK. 
The first two terms on the right-hand side of Eq.~(\ref{eq1}) are characteristic of a degenerate 2D Fermi liquid with spin fluctuations \cite{Ogura(1997)}. 
$C_{\text{amor}}(T)$ is the heat capacity of the 0.45~nm$^{-2}$ sample. 
This is associated with nuclear spin degrees of freedom of amorphous $^3$He \cite{Golov(1996)} trapped on strong adsorption sites of Grafoil. 
The unusually weak $T$-dependence is a result of a wide distribution of exchange interaction in the amorphous state. 
As shown in Fig. \ref{lfig_1}(b), with increasing $\rho$, only the amorphous component (fitted $\beta$ value) increases linearly with a small offset of 0.1~nm$^{-2}$, and is saturated above ${\rho}$ ${\approx}$ 0.6~nm$^{-2}$. 
After then, the fitted $\gamma$ value starts to increase linearly 
until a kink at $\rho$ = 1.4~nm$^{-2}$ near $\gamma$ = $\gamma_{\text{ideal}}$. 
The density variation of $\gamma$ above 0.6~nm$^{-2}$ is very similar to that observed in the third layer above the intervening region~\cite{Sato(2010)}. 
To our knowledge, the only reasonable explanation for this is the phase separation between a degenerate Fermi liquid ($puddles$) with an almost fixed density of $\rho_{c0}$ (= 0.8~nm$^{-2}$) and a dilute gas phase with a negligibly small $C$ contribution (see later discussion). 
It is clear that the G-L transition in the first layer develops on the uniform region of the substrate independently of the preceding occupation of the heterogeneous sites by $^3$He. 
Such sites would be located only near platelet edges. 
The number of $^3$He atoms contributing to $C_{\text{amor}}$ is 10\% of that on the uniform surface to complete the ${\sqrt{3}} {\times} {\sqrt{3}}$ commensurate phase \cite{RhoCorrection}.
This ratio is consistent with previous thermodynamic measurements \cite{Elgin(1978)}.

\begin{figure}[b]
\includegraphics[width=0.46\textwidth]{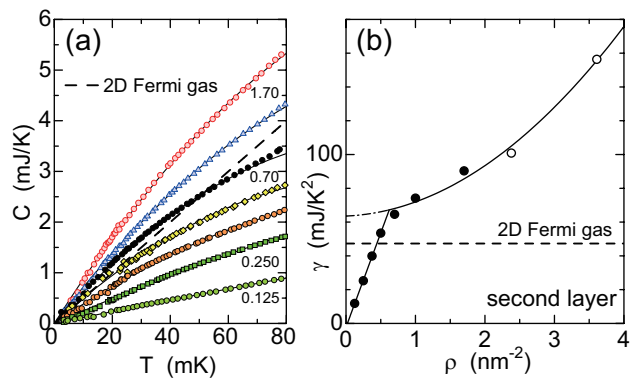}
\caption{
(a) Heat capacities of the second layer of $^3$He on Grafoil preplated with a monolayer of $^4$He. 
The solid lines are fittings to Eq. (\ref{eq2}).
(b) Density variation of the fitting parameter $\gamma$ in Eq. (\ref{eq2}). 
The open circles are from Matsumoto {\it et al.}~\cite{Matsumoto(2005)}.
Note that their data point at 2.38~nm$^{-2}$ is not shown in Ref.~\cite{Matsumoto(2005)}.
Otherwise, notation here is the same as in Fig.~\ref{lfig_1}.
}
\label{lfig_2}
\end{figure}

Next, we made heat capacity measurements of the second layer of $^3$He on Grafoil preplated with a monolayer of $^4$He, which preferentially occupies the first layer because of its smaller zero-point energy than $^3$He. 
This technique has widely been employed in previous experiments \cite{Sato(2010), Lusher(1991), Collin(2001)}. 
We introduced exactly the same amount of $^4$He (12.09~nm$^{-2}$) as that in Ref.~\cite{Sato(2010)}. 
As seen in Fig. \ref{lfig_2}(a), any non-Fermi liquid $C$ contributions are absent here, indicating thorough elimination of the substrate heterogeneity effect by the $^4$He preplating.
The data can be fitted to the formula:
\begin{equation}\label{eq2}
   C(T) = {\gamma} T - {\alpha} T^{2}
\end{equation}
very well. 
The fitted $\gamma$ follows perfectly the $\rho$-linear dependence with a negligibly small offset (0.02~nm$^{-2}$) as well as a kink at $\rho_{c0}$ = 0.6~nm$^{-2}$ and $\gamma_{c0}$ = $1.3\gamma_{\text{ideal}}$ (see Fig. \ref{lfig_2}(b)). 
Therefore, a G-L transition is observed again. 
Moreover, the second layer of $^3$He should be the best representative of monolayer $^3$He on graphite without heterogeneities.
According to the previous experimental  \cite{Bretz(1973)} and theoretical \cite{Pierce(2000)} determinations of the second layer promotion density of $^4$He ($11.4\sim11.8$~nm$^{-2}$), we expect that a small fraction ($0.3\sim0.7$~nm$^{-2}$) of $^4$He is promoted to the second layer and preferentially occupies deeper potential sites above the substrate heterogeneities. 
This explains why we don't observe $C_{\text{amor}}$ nor a sizable intervening region prior to the puddle region in the second-layer measurement. 
Our $\gamma$ data follow smoothly the previous data \cite{Matsumoto(2005)} using exactly the same experimental setup (open circles in Fig. \ref{lfig_2}(b)) at $\rho \ge \rho_{c0}$, where $\gamma$, and hence the quasiparticle effective mass, increases progressively due to particle correlations.

Let us briefly comment on possible finite-size effects caused by the platelet structure of Grafoil. 
The energy discreteness estimated from the platelet size is $2\sim200$~$\mu$K. 
This will not affect at least the leading $\gamma$$T$-terms in Eqs. (\ref{eq1}) and (\ref{eq2}), and hence our G-L transition scenario, within the temperature range we studied ($T\ge$ 2~mK). 
On the other hand, the correction term $\alpha T^{2}$, which is due to the spin fluctuations 
\cite{Ogura(1997)}, is suppressed depending on the puddle size within the two-phase coexistence region of the second layer. 
Eventually, $\alpha/\gamma$ decreases from 5 K$^{-1}$ to zero with decreasing $\rho$ presumably because of the long-wavelength cutoff of the fluctuations.
More details of the size effects will be discussed elsewhere \cite{Sato(QFS2012)}.

\begin{figure}[h]
\includegraphics[scale=0.37]{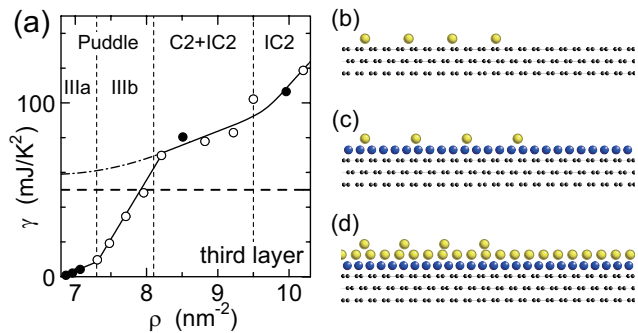}
\caption{
(a) Density variation of $\gamma$ deduced from measured heat capacities of the third-layer liquid $^3$He on graphite preplated with a monolayer $^4$He; present work (closed circles), Ref.~\cite{Sato(2010)} (open circles).
The magnetic contribution from the second-layer solid $^3$He has already been subtracted as described in Ref.~\cite{Sato(2010)}. 
The solid line is guide for eye. 
The uniform liquid region above 8.1~nm$^{-2}$ is divided into two regions, (C2 + IC2) and IC2, depending on the structure of the second-layer solid $^3$He (see Ref.~\cite{Sato(2010)}).
Schematic cross-sectional views of the first (b), second (c) and third-layer (d) $^3$He puddles on graphite. Only topmost three graphene-layers are drawn here.
}
\label{lfig_3}
\end{figure}

We have made additional heat-capacity measurements for the third layer of $^3$He to understand further details of the density variation of $\gamma$ studied in Ref.~\cite{Sato(2010)}. 
Within the intervening region between 6.8 and 7.3~nm$^{-2}$ (Region IIIa in Fig. \ref{lfig_3}(a)), it was found that $\gamma$ varies in proportion to $\rho$ with a factor of three smaller slope than that in the following main puddle region  of 7.3~$\le \rho \le$ 8.1~nm$^{-2}$ (Region IIIb). 
We speculate that, in Region IIIa, promotion to the third layer as liquid puddles and compression of the second layer proceed simultaneously. 
This speculation is supported by the observed increases of magnetic $C$-isotherms below 1~mK by about 10\% in the corresponding density region \cite{Sato(2010)}. 
A similar intervening region is also observed in the previous NMR experiment \cite{Collin(2001)}, where the second layer of $^3$He is compressed by adding $^4$He. 
The compression of the second layer can either be solidification of a remnant high-density liquid, which may exist  nearby the heterogeneities, or introduction of iterstitial atoms to the commensurate phase (C2)~\cite{Takagi(private_comm)}.
Consequently, we estimate $\rho_{c0}$ in the third layer as 0.9~nm$^{-2}$ or slightly less.
In Figs. \ref{lfig_3}(b), (c) and (d), cross-sectional views of the first, second and third-layer puddles of $^3$He are imaged, respectively. 

\begin{figure}[h]
\includegraphics[scale=0.27]{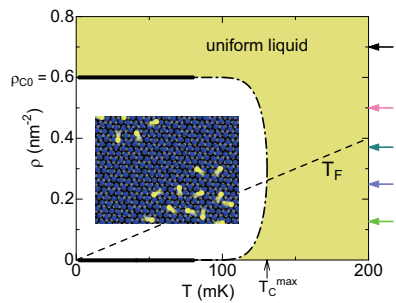}
\caption{
Low-density phase diagram of monolayer $^3$He on graphite.
The thick solid lines are the gas-liquid (G-L) transition lines determined from this experiment.
The dash-dotted line is a calculated one normalized to $\rho_{c0}$ = 0.6~nm$^{-2}$ and  ${T_c}^{max}$ = 130~mK (Ref. \cite{Ostlund(1980)}).
The dashed line represents $T_{F}$ of 2D $^3$He gas.
The arrows denote sample densities at which we made the $C$ measurements for the second layer of $^3$He. 
The inset shows a schematic top view of phase-separated liquid puddles in the second layer.
}
\label{lfig_4}
\end{figure}

The zinc superconducting heat-switch we used unfortunately does not allow us to extend our heat-capacity measurements beyond 80 mK where one expects to observe $C$ anomalies associated with finite-$T$ G-L transitions (${T_c}$). 
We speculate that the highest ${T_c}$ (${T_c}^{max}$) is realized at $\rho \approx \rho_{c0}$/2 in a $T$-range betwen 80~mK and 0.7~K.
The high-$T$ bound comes from the known ${T_c}^{max}$ for $^4$He in 2D \cite{Greywall(1993)}. 
Then, a naive question is why we don't observe any $C$ contributions from the phase-separated gas phase in the puddle regions. 
The dash-dotted line in Fig. \ref{lfig_4} is a G-L phase separation line calculated for classical adatoms interacting with the Lennard-Jones potential \cite{Ostlund(1980)}, where we adjusted the line so as to give $\rho_{c0}=$ 0.6~nm$^{-2}$ and ${T_c}^{max}$ = 130~mK. 
 This ${T_c}^{max}$ value was chosen arbitrarily.
The low-density branch and high-density one give the equilibrium gas density  ${\rho_c}^{g}$~($T$) and liquid one ${\rho_c}^{l}$~($T$), respectively. 
Since they vary exponentially with $T$ at $T<{T_c}^{max}$, we expect ${\rho_c}^{g}$/$\rho_{c0} \le$~0.02 and ${\rho_c}^{l}$/$\rho_{c0}\approx$ 1 at $T\le$~80~mK. 
That is why the $C$ contribution from the gas phase is immeasurably small and the liquid phase is always degenerate with nearly the constant density $\rho_{c0}$ in our measurement.  
If the G-L phase separation does not occur, we should observe a smooth approach of $\gamma$ to $\gamma_{\text{ideal}}$ when $\rho$ decreases down to the lowest density sample (0.125~nm$^{-2}$; Fermi temperature ($T_{F}$) = 63~mK) without any kinks, which is of course  totally not the experimental case.

The fact that the first three layers of $^3$He on graphite have nearly the same $\rho_{c0}$ values (= 0.8, 0.6, 0.9~nm$^{-2}$) excludes the quasi two-dimensionality from possible explanations for the self-binding. 
This is because the confinement potentials and wave-function overlappings between the successive layers are quite different each other in these monolayers. 
Furthermore, the indirect interaction ($V_{ind}$) mediated by excitations in the underlayer should be quite different, too. 
Schick and Campbell \cite{Schick(1970)} calculated $V_{ind}$ due to phonon exchange to be proportional to a factor $n_sc_T^{-2}\epsilon^2$ in the case of substrates occupying a half-infinite space. 
Here $n_s$, $c_T$ and $\epsilon$ are the three-dimensional density and phonon velocity of the substrate and  the minimum of the He-substrate potential, respectively. 
If we apply this theory to the present problem, this factor is, at least, an order of magnitude smaller for the first layer compared to the third layer.
This is inconsistent with the fact that nearly the same $\rho_{c0}$ values are obtained in these layers~\cite{Cole(1981)}. 
We thus conclude that the observed G-L transition in the present experiment should be an intrinsic property of $^3$He in strictly 2D. 

Our conclusion gives rise to a conflict with the existing many-body calculations for $^3$He in 2D \citep{Novaco(1975),Miller(1978),Krishnamachari(1999),Grau(2002),Um(1997)}.  
It is, however, worthwhile to remark on recent variational and diffusion Monte Carlo calculations by Kili{\'c} and Vranje{\v{s}}~\cite{Kilic(2004)} on binding energies of $^3$He molecules of $N$ atoms in 2D. 
They obtained tiny but finite binding energies, $-(0.02$$\sim$$0.04)$~mK, for 2~$\le$$N$$\le$~6. 
Since the binding energy should decrease with $N\rightarrow\infty$, their calculations seem to be consistent with the present experimental result.

In summary, we found the gas-liquid transition in the three different $^3$He monolayer systems, i.e., the first, second and third layers, on graphite from the heat capacity measurements at low densities never explored before in the degenerate temperature region down to 2 mK. 
The phase-separated liquid phases have surprisingly similar densities ($\rho_{c0}= 0.6\sim0.9$~nm$^{-2}$) despite their quite different environments, which indicates that $^3$He atoms in a strictly 2D space are self-bound forming liquid puddles at the ground state. 
The mean interatomic distance in this puddle is very large (1.1$\sim$1.4~nm). 
This would be to our knowledge the lowest density liquid ever found in nature. 
The present result contradicts the existing many-body calculations for $^3$He in 2D, providing an important constraint on theory. 
In future, it will be highly desirable to detect directly the expected thermodynamic anomalies at the critical temperature. 

We thank Tony Leggett for illuminating discussions and suggestions. 
This work was financially supported by Grant-in-Aid for Scientific Research on Priority Areas (No. 17071002) from MEXT, Japan and Scientific Research (A) (No. 22244042) from JSPS. 
DS acknowledges the support from JSPS Research program for Young Scientists.


\begin{thebibliography}{99}
\bibitem{Note}Note that laser-cooled alkali atoms stay gaseous down to the nanokelvin 
 region as a metastable state only at very low densities where three-body 
 collisions are not significant.
\bibitem{RhoCorrection}The $^3$He density on the uniform region ($\rho'$) can be different from the nominal density ($\rho$) calibrated only at a single density, e.g., the density at the $\sqrt{3}\times\sqrt{3}$ commensurate phase in the first layer (6.37~nm$^{-2}$). 
For our Grafoil substrate, the relation between $\rho'$ and $\rho$ is $\rho'(\text{nm}^{-2})= 1.10 (\rho - 0.6)$ at least for the 1st layer.
\bibitem{Cole(1981)}If we extrapolate the smooth density variations of $\gamma$ above the kinks to zero density, we have $(1.2\sim1.3)\gamma_{\text{ideal}}$ for the three monolayers.
This is indicative of mass enhancement of a single $^3$He atom caused by interactions with the underlayers.
If so, $V_{ind}$ may play certain roles in the observed self-binding.
Further theoretical studies in this aspect are desirable.
\end{thebibliography}
\end{document}